\documentstyle[twoside,fleqn,espcrc2,epsfig]{article}
\newcommand{\be}{\begin{equation}}
\newcommand{\ee}{\end{equation}}
\newcommand{\ba}{\begin{eqnarray}}
\newcommand{\ea}{\end{eqnarray}}

\makeatletter
\newbox\slashbox \setbox\slashbox=\hbox{\large$/$}
\def\pslash#1{\setbox\@tempboxa=\hbox{$#1$}
  \@tempdima=0.5\wd\slashbox \advance\@tempdima 0.5\wd\@tempboxa
  \copy\slashbox \kern-\@tempdima \box\@tempboxa}
\def\FMSlash{\protect\pslash}
\title{Lowest eigenvalues of the Dirac operator for two color
QCD at nonzero chemical potential}
\author{Elmar Bittner\address{Atominstitut, Technische Universit\"at Wien,
        A-1040 Vienna, Austria}, Maria-Paola Lombardo\address{Istituto Nazionale di Fisica
        Nucleare, Sezione di Padova, e Gruppo Collegato di Trento, Italy},
        Harald Markum$^{\rm a}$, and Rainer Pullirsch$^{\rm a}$}
\begin{document}
\begin{abstract} 
  We investigate the eigenvalue spectrum of the staggered Dirac matrix
  in SU(3) and U(1) gauge theory as well as in full QCD with two colors
  and finite chemical potential.
  Along the strong-coupling axis up to the phase transition, 
  the low-lying Dirac spectrum of these quantum field
  theories is well described by random matrix theory and exhibits universal
  behavior. Related results for gauge theories with minimal coupling are 
  discussed in the chirally symmetric phase and no universality is seen
  for the microscopic spectral densities.
\end{abstract}
\date{\today}
\maketitle

\section{Spectrum in confinement}

The eigenvalues of the Dirac operator are of great
interest for the universality of important features of QCD and QED.
The accumulation of small eigenvalues is, via the
Banks-Casher formula~\cite{Bank80}, related to the spontaneous
breaking of chiral symmetry. Their
properties are known to be described by random matrix theory (RMT)
in the confinement, see Ref.~\cite{Berb98}.

The Dirac operator itself, $\FMSlash{D}=\FMSlash{\partial}+ig\FMSlash{A}$, is
anti-Hermitian so that the eigenvalues $\lambda_n$ of $i\FMSlash{D}$
are real.  Because of $\{\FMSlash{D},\gamma_5\}=0$ the nonzero
$\lambda_n$ occur in pairs of opposite sign.  
In the presence of a chemical potential $\mu > 0$, the 
Euclidean action becomes
a complex number and the fermionic matrix becomes a non-Hermitian matrix.

%%%%%%%%%%%%%%%%%%%%%%%%%%%%%%%%%%%%%%%%%%%%%%%%%%%%%%%%%%%%%%%%%%%%%%%%%%%%%%%%
\section{Spectrum into deconfinement}

\begin{figure*}[p]
    \hspace*{3mm}\centerline{\psfig{figure=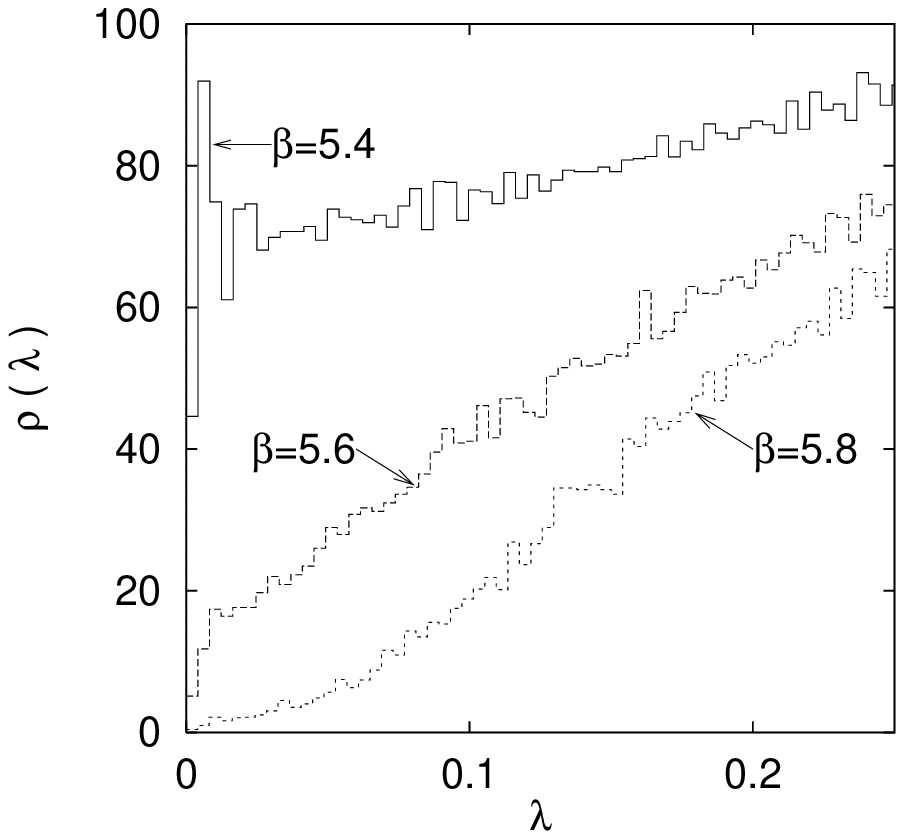,width=5cm}\hspace*{2mm}
                \psfig{figure=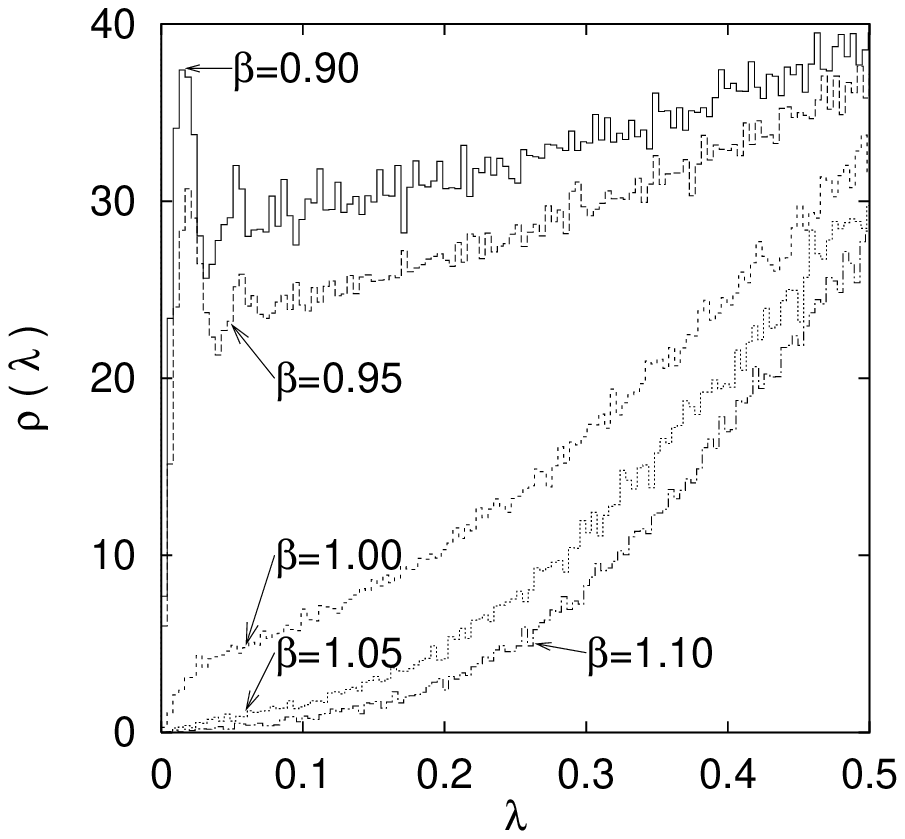,width=5cm}\hspace*{2mm}
                \psfig{figure=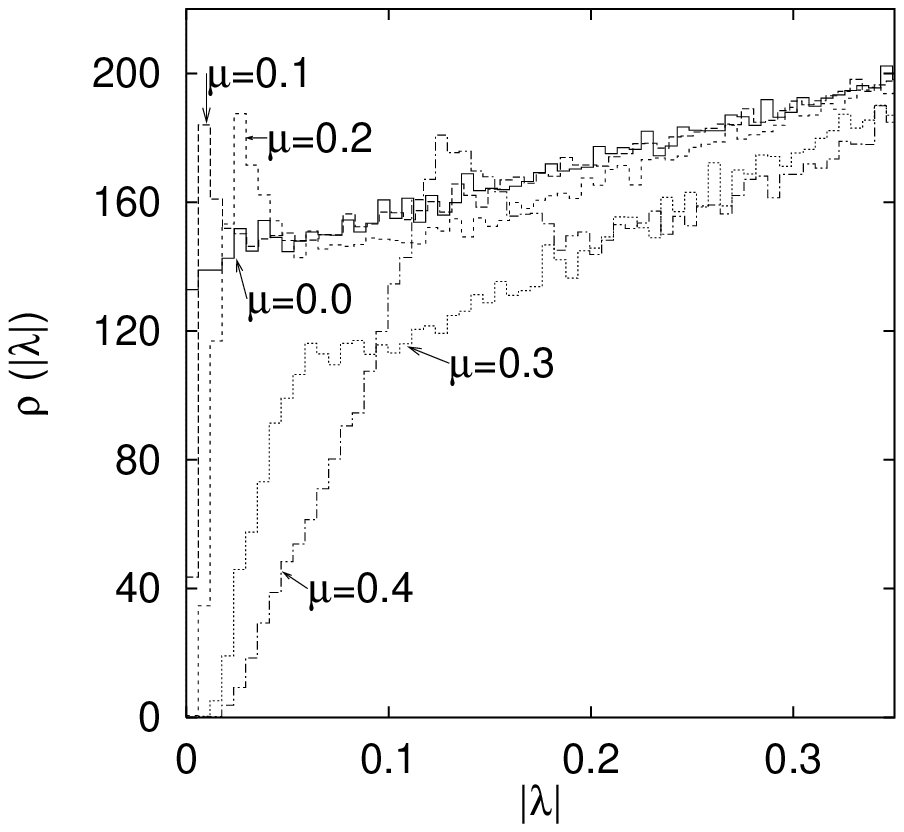,width=5cm}}
\vspace*{-10mm}
  \caption{Density $\rho (\lambda)$ of small eigenvalues for SU(3) (left) and
           U(1) gauge theory (center) as well as for two-color QCD (right)
           across the transition of critical coupling and critical chemical
           potential, respectively.}
  \label{fig4}
\vspace*{5mm}
  \hspace*{3mm}\centerline{\psfig{figure=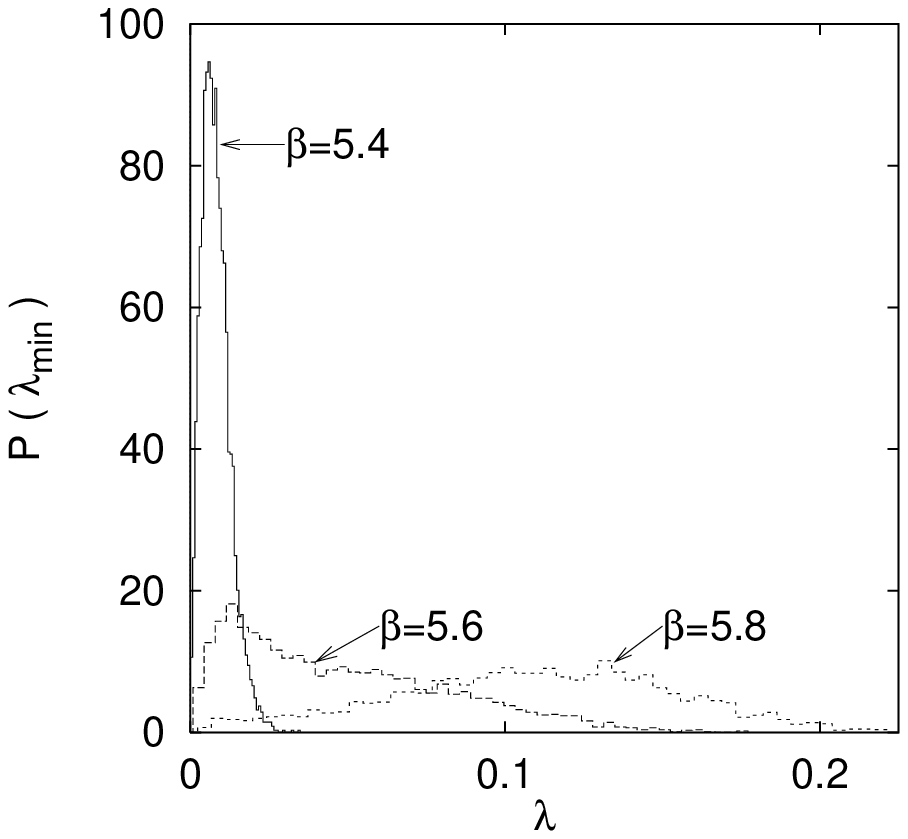,width=5cm}\hspace*{2mm}
  \psfig{figure=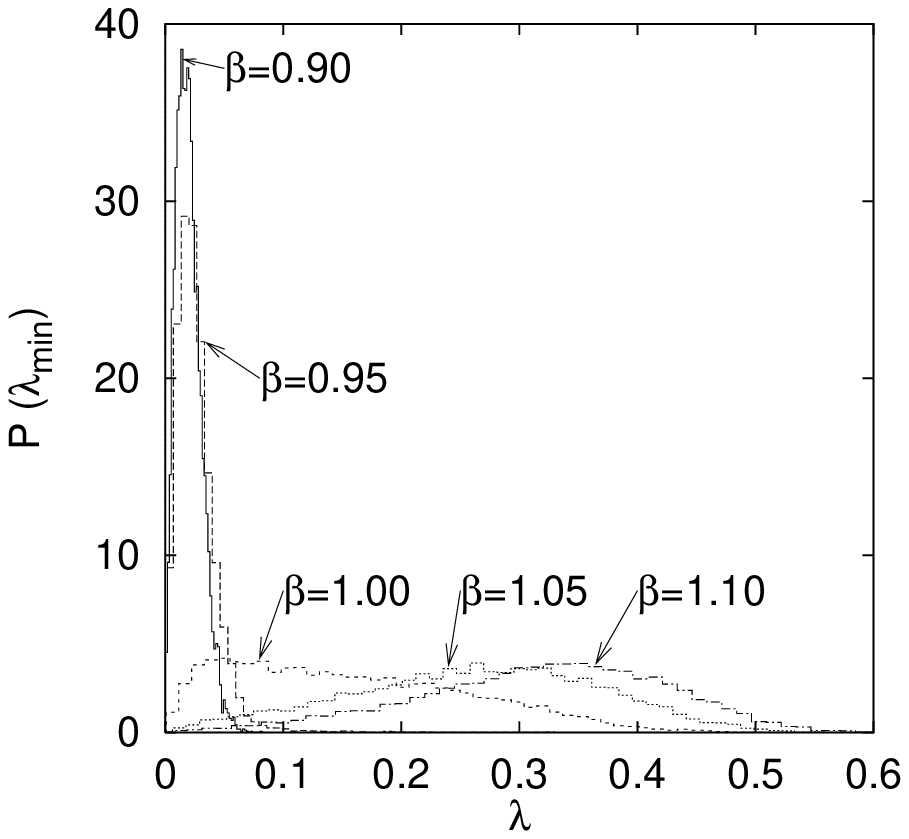,width=5cm}\hspace*{2mm}
  \psfig{figure=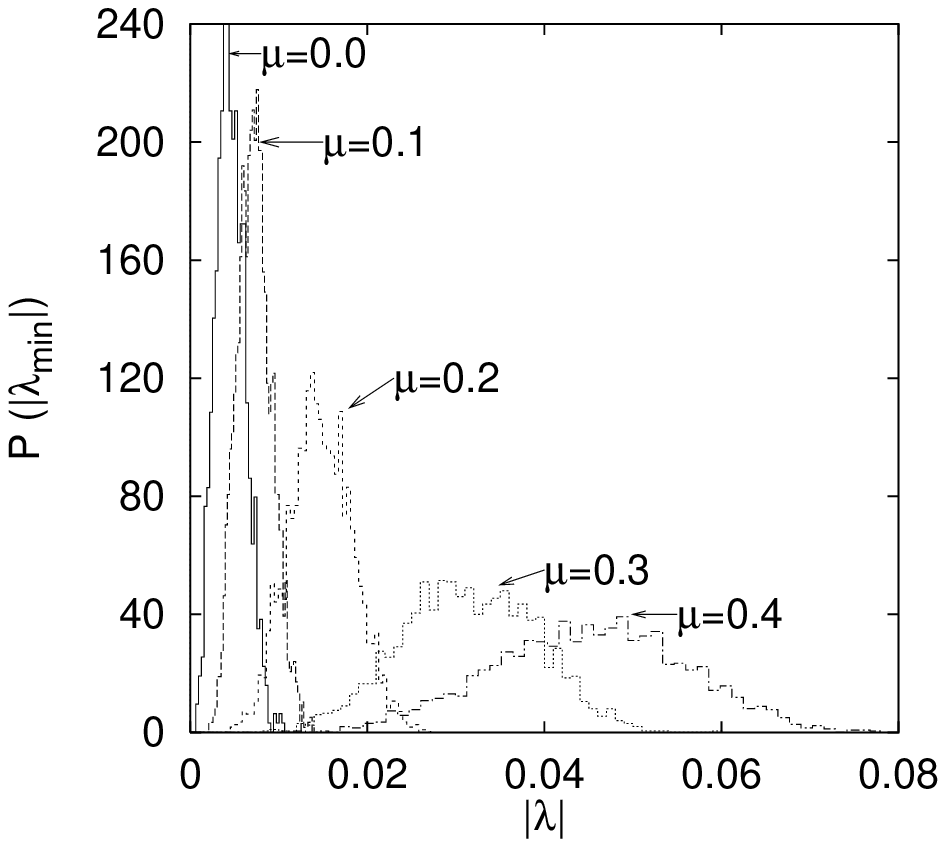,width=5cm}}
\vspace*{-10mm}
  \caption{Distribution $P(\lambda_{\min})$ of the smallest eigenvalue as in 
           Fig.~\ref{fig4}.}
  \label{fig5}
\vspace*{5mm}
  \hspace*{3mm}\centerline{\psfig{figure=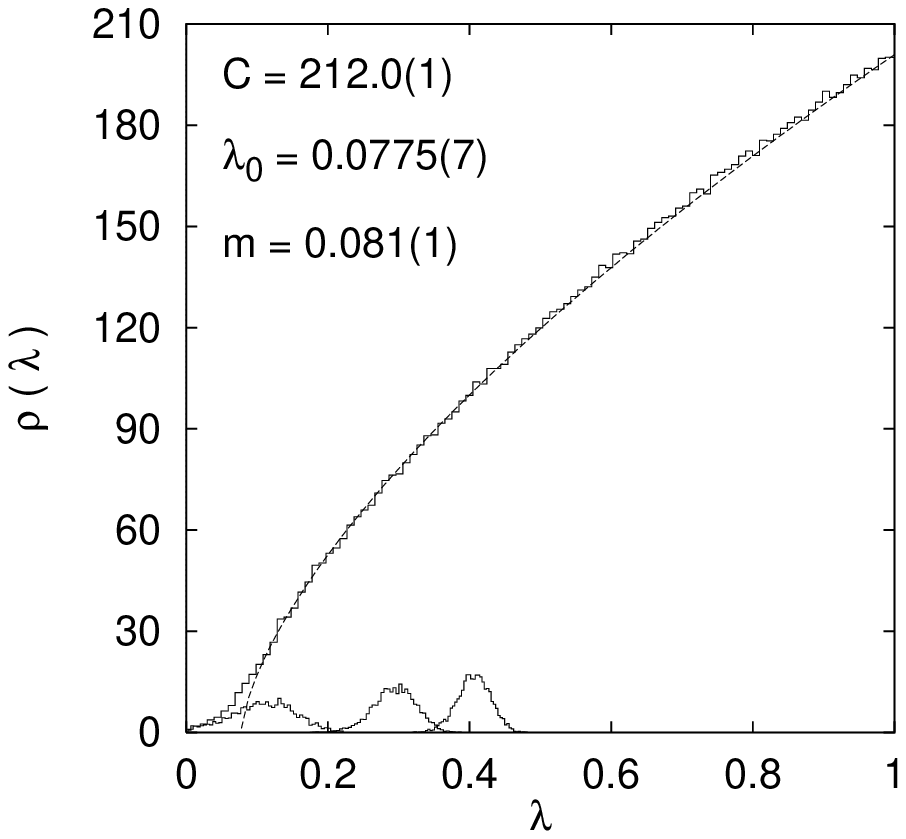,width=5cm}\hspace*{2mm}
  \psfig{figure=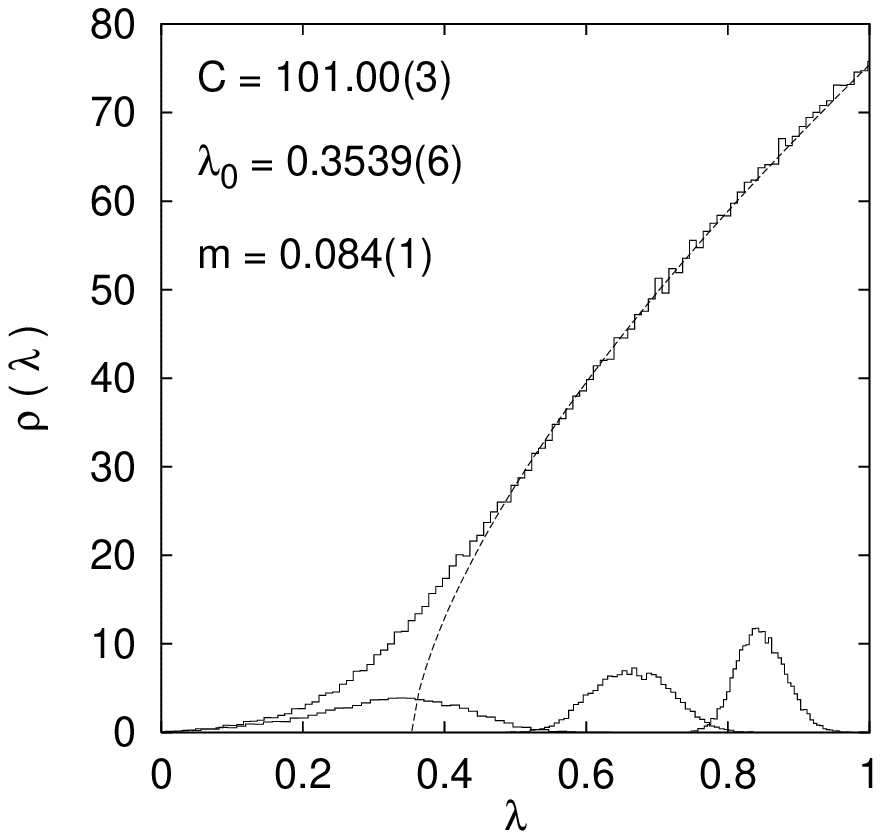,width=5cm}\hspace*{2mm}
  \psfig{figure=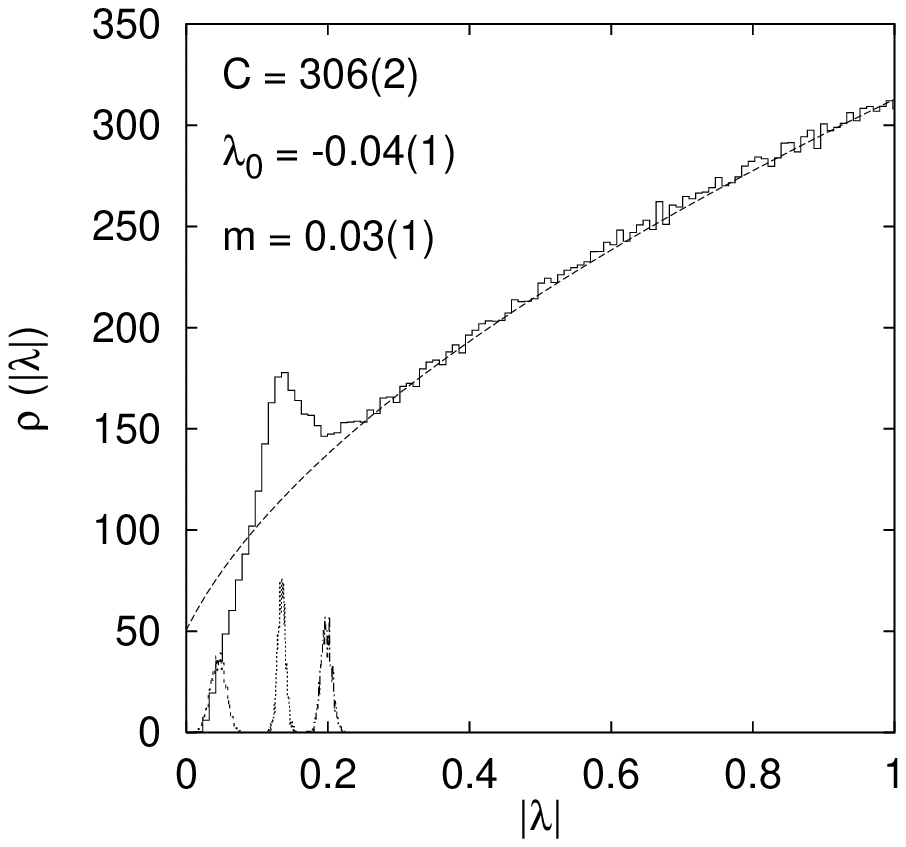,width=5cm}}
\vspace*{-10mm}
  \caption{Fit of the spectral density to $\rho (\lambda) =
    C (\lambda -\lambda_0)^{2m+1/2}$ in SU(3) at $\beta=5.8$ (left),
    in U(1) at $\beta=1.10$ (center) and in two-color QCD at
    $\mu=0.4$ (right). The contribution
    of the smallest eigenvalue, the $11^{{\mbox{th}}}$ eigenvalue and the
    $21^{{\mbox{st}}}$ eigenvalue is inserted.}
  \label{fig6}
\end{figure*}

We have continued our investigations~\cite{Bitt00} with a study of the
distribution of the small eigenvalues in the whole phase diagram. The
Banks-Casher formula \cite{Bank80} relates the Dirac eigenvalue density
$\rho(\lambda)$ at $\lambda=0$ to the chiral condensate,
$ \Sigma \equiv |\langle \bar{\psi} \psi \rangle| =
 \lim_{\varepsilon\to 0}\lim_{V\to\infty} \pi\rho (\varepsilon)/V$.
The microscopic spectral density,
$ \rho_s (z) = \lim_{V\to\infty}
 \rho \left( {z/V\Sigma } \right)/V\Sigma , $
should be given by the appropriate prediction of RMT~\cite{ShVe92}.
In the case of complex eigenvalues the situation is more complicated
\cite{Spli01} and $\rho(0)$ can be used as a lower bound for 
$\Sigma$.

We present results for the density of the small eigenvalues in 
Fig.~\ref{fig4} for SU(3) theory and the
staggered Dirac operator on a $4^4$ lattice from 5000 configurations
for $\beta=5.4$ and 3000 configurations for $\beta=5.6$ and $\beta=5.8$
around the critical temperature $\beta_c \approx 5.7$.
Our analogous presentation for U(1) theory is from 10000 configurations on a $4^4$
lattice around the critical coupling $\beta_c \approx 1.01$.
In the confinement
phase it was shown that, both the microscopic spectral density $\rho_s(z)$ and the 
distribution $P(\lambda_{\rm min})$ of the smallest eigenvalue
agree with the RMT predictions of the chiral unitary ensemble
for topological charge $\nu = 0$ \cite{Goec99,Berb98}.
In the case of two-color QCD with staggered fermions on a $6^4$
lattice we produced at least 2100 configurations for each value
of $\mu$ around the critical chemical potential $\mu_c \approx 0.3$
keeping $\beta =1.3$ fixed~\cite{Hand99}. Since the eigenvalues
move into the complex plane for $\mu > 0$, a band of width $\epsilon
= 0.015$ parallel to the imaginary axis was considered to construct
$\rho(y)$, i.e. $\rho(y)\equiv\int_{-\epsilon}^\epsilon
dx\,\rho(x,y)$, where $\rho(x,y)$ is the density of the complex
eigenvalues $x+iy$. Alternatively $\rho(|\lambda|)$ was constructed
from the absolute value for $|\lambda|$ small and is presented in this
write-up.

The distribution of the lowest eigenvalue is displayed in Fig.~\ref{fig5}
for the theories under investigation. The data of the SU(3) and U(1)
theories in the deconfinement could not be fitted to the width of the 
functional form derived for RMT in the chirally broken phase. The quality
of the data of two-color QCD was not sufficient for a reliable fit to
a trial function for $P(|\lambda_{\min}|)$
except for the RMT prediction of the chiral symplectic
ensemble at $\mu=0$.
The lattice results of the three theories look very similar and might give some
help for a derivation within RMT. 

Nevertheless, the quasi-zero modes which are responsible for the chiral
condensate $\Sigma \neq 0$ build up when we cross from the deconfinement
into the confined phase.
Figures~\ref{fig4} and~\ref{fig5} demonstrate that both $\rho(\lambda)$
and $P(\lambda_{\min})$ plotted with varying $\beta$ or $\mu$ on identical scales,
respectively, can serve as an indicator for the phase transition.
The change in bending from positive to negative curvature of
$\rho(0)$ as a function of coupling/density might serve to 
pin down the critical point.

In Fig.~\ref{fig6} we turn to a discussion of the spectrum in the 
quark-gluon plasma and Coulomb phase. From RMT a functional form of
$\rho (\lambda) = C (\lambda -\lambda_0)^{2m+1/2}$ is expected at
the onset of the eigenvalue density~\cite{Bowi91}. A fit to the data
in the regime up to $\lambda = 1$ yields
$m=0.081(1)$ for SU(3) and $m=0.084(1)$ for the Abelian theory,
in agreement with recent studies~\cite{Lang99}. This suggests that
both theories correspond to universality class $m=0$. For this class
a microscopic level density involving the Airy function can be deduced
from RMT~\cite{Forr93}. A rescaling of our data from the $4^4$ lattice
to this functional form is not satisfactory for both theories~\cite{Lang99}.
We tried a fit for two-color QCD assuming the above functional
form for $\rho(|\lambda|)$ in a region up to $|\lambda|=1$ obtaining
$m=0.03(1)$ consistent with universality class $m=0$.

%%%%%%%%%%%%%%%%%%%%%%%%%%%%%%%%%%%%%%%%%%%%%%%%%%%%%%%%%%%%%%%%%%%%%%%%%%%%%%%%
\section{Conclusion}

We investigated universality concerning the low-lying spectra of the
Dirac operators of both SU(3) and U(1) gauge theories. One
finds that in the phase in which  chiral symmetry is spontaneously
broken the distribution
$P(\lambda_{\min})$ and the microscopic spectral density $\rho_s(z)$
are described by chiral RMT. 
For two-color QCD with chemical potential we were not able to
verify such a relation for $0<\mu<\mu_c$, not only because of our data but 
also in lack of an analytic result for non-Hermitian RMT~\cite{akem}.
In the phase where chiral symmetry is restored one has to rely on ordinary RMT.
Here we find universal behavior only of the macroscopic density $\rho(\lambda)$
for all gauge theories with minimal coupling.

{\it Acknowledgments:}
This study was supported in part by FWF project P14435-TPH.

%%%%%%%%%%%%%%%%%%%%%%%%%%%%%%%%%%%%%%%%%%%%%%%%%%%%%%%%%%%%%%%%%%%%%%%%%%%%%%%%


\begin{thebibliography}{99}
\bibitem{Bank80} T. Banks and A. Casher, Nucl. Phys. B 169 (1980) 103.
\bibitem{Berb98} M.E. Berbenni-Bitsch, S. Meyer, A. Sch\"afer,
J.J.M. Verbaarschot, and T. Wettig, Phys. Rev. Lett. 80 (1998) 1146;
B.A. Berg, H. Markum, R. Pullirsch, and T. Wettig, Phys. Rev. D 
63 (2001) 014504.
\bibitem{Bitt00} E. Bittner, H. Markum, and R. Pullirsch,
Nucl. Phys. B (Proc. Suppl.) 96 (2001) 189; 
E. Bittner, M.-P. Lombardo, H. Markum, and R. Pullirsch,
Nucl. Phys. B (Proc. Suppl.) 94 (2001) 445.
\bibitem{ShVe92} E.V. Shuryak and J.J.M. Verbaarschot, Nucl. Phys.
  A 560 (1992) 306; J.J.M. Verbaarschot and I. Zahed, Phys. Rev. Lett.
  70 (1993) 3852; J.J.M. Verbaarschot, Phys. Rev. Lett. 72 (1994) 2531.
\bibitem{Spli01} K. Splittorff, D. Toublan, and J.J.M. Verbaarschot,
hep-ph/0108040.
\bibitem{Goec99} M. G\"ockeler, H. Hehl, P.E.L. Rakow, A. Sch\"afer, and
 T. Wettig, Phys. Rev. D 59 (1999) 094503.
\bibitem{Hand99} S. Hands, J.B. Kogut, M.-P. Lombardo, and
  S.E. Morrison, Nucl. Phys. B 558 (1999) 327.
\bibitem{Bowi91} M.J. Bowick and E. Br\'ezin, Phys. Lett. B 268 (1991) 21;
 E. Kanzieper and V. Freilikher, Phys. Rev. E 55 (1997) 3712.
\bibitem{Lang99} F. Farchioni, P. de Forcrand, I. Hip, C.B. Lang, and
 K. Splittorff, Phys. Rev. D 62 (2000) 014503;
 P.H. Damgaard, U.M. Heller, R. Niclasen, and K. Rummukainen, Nucl. Phys.
 B 583 (2000) 347.
\bibitem{Forr93} P.J. Forrester, Nucl. Phys. B 402 (1993) 709.
\bibitem{akem}  G. Akemann, hep-th/0106053.
%%%%%%%%%%%%%%%%%%%%%%%%%%%%%%%%%%%%%%%%%%%%%%%%%%%%%%%%%%%%%%%%%%%%%%%%%%%%%%%%

\end{thebibliography}
\end{document}